\documentclass[a4paper,11pt]{article}
\usepackage{pos}

\title{Higher-order corrections to top-quark pair production in the SMEFT}

\renewcommand{\printHeadAuthors}{A. Tonero and N. Kidonakis}

\author*{Alberto Tonero}
\author{Nikolaos Kidonakis}

\affiliation{Department of Physics, Kennesaw State University, \\
Kennesaw, GA 30144, USA}

\emailAdd{atonero@kennesaw.edu}
\emailAdd{nkidonak@kennesaw.edu}

\abstract{We present theoretical results at approximate NNLO in QCD for top-quark pair-production total cross sections and top-quark differential distributions at the LHC in the SMEFT. In this work we consider effects of new physics parametrized by the chromomagnetic dipole operator. These approximate results are obtained by adding higher-order soft gluon corrections to the complete NLO calculations. The higher-order corrections are large, and they reduce the scale uncertainties. These improved theoretical predictions can be used to set stronger bounds on top-quark QCD anomalous couplings. }

\FullConference{42nd International Conference on High Energy Physics (ICHEP2024)\\
18-24 July 2024\\
Prague, Czech Republic\\}

\begin{document}
\maketitle

\section{Introduction}
The top quark is the heaviest elementary particle in the Standard Model (SM), and it couples most strongly to the electroweak symmetry breaking sector. The top-quark production cross section at the Large Hadron Collider (LHC) is large and, thus, precision measurements can be performed with potential sensitivity to beyond the Standard Model (BSM) physics such as Standard Model Effective Field Theory (SMEFT)~\cite{Weinberg:1979sa,Buchmuller:1985jz,Grzadkowski:2010es}. In this model, new physics is parametrized by a set of higher-dimensional operators organized in a series expansion with increasing operator dimension.
To extract consistent results, the theoretical predictions need to match the precision of the experimental measurements, which has increased after LHC Run II. Calculations of NLO QCD corrections in the SMEFT can be done in {\small \sc MadGraph5\_aMC@NLO}~\cite{MG5} through the {\small \sc SMEFT@NLO} package~\cite{Degrande:2020evl}.

In this talk, we present the results of Ref.~\cite{Kidonakis:2023htm} where we produced more accurate theoretical predictions for top-quark pair production at the LHC in the SMEFT by focusing on the so-called chromomagnetic dipole operator.  We improved the NLO theoretical estimate~\cite{BuarqueFranzosi:2015jrv} by computing the total cross sections and top-quark differential distributions in $t{\bar t}$ production at approximate next-to-next-to-leading order (aNNLO) in QCD. These results are derived by adding second-order soft-gluon corrections (for a review see~\cite{NKtoprev}) to the complete NLO QCD calculation. 

\section{SMEFT contributions to $t{\bar t}$ production}
Among the possible dimension-six operators that  belong to the SMEFT Lagrangian~\cite{Grzadkowski:2010es}, we focus on the so-called chromomagnetic dipole operator which is frequently considered in recent experimental analyses on SMEFT in $t\bar t$ production~\cite{CMS:2018adi,CMS:2019zct,ATLAS:2022xfj,ATLAS:2022mlu}. The chromomagnetic dipole operator is defined as 
\begin{equation}\label{eftop}
\frac{c_{tG}}{\Lambda^2}{\cal{O}}_{tG}+{\rm h.c.}
\end{equation}
where ${\cal{O}}_{tG}=g_S \bar q_{3L} \sigma_{\mu\nu}T^A t_R\tilde \varphi G_A^{\mu\nu}$
with $g_S$ the strong coupling, $T^A$ the $SU(3)$ generators, $\sigma_{\mu\nu}=i[\gamma_{\mu},\gamma_{\nu}]/2$ with $\gamma_{\mu}$ the Dirac matrices, $q_{3L}=(t\quad b)_L$, $\varphi$ the Higgs doublet, and $G_{\mu\nu}^A$ the gluon field strength tensor. The $c_{tG}$ coefficient is dimensionless and taken to be real. 
As in the SM, top-antitop production in the presence of the chromomagnetic dipole operator can occur at leading order (LO) through the $q{\bar q} \to t{\bar t}$ and $gg \to t{\bar t}$ channels. We write   
\begin{equation}
f_{1}(p_1)\, + \, f_{2}\, (p_2) \rightarrow t(p_t) \, + \, {\bar t}(p_{\bar t})  \, ,
\label{processes}
\end{equation}
where $f_1$ and $f_2$ are quarks or gluons, and we define $s=(p_1+p_2)^2$, $t=(p_1-p_t)^2$, $u=(p_2-p_t)^2$, and $s_4=s+t+u-2m_t^2$ with $m_t$ the top quark mass. The partonic differential cross section at LO can be written as
\begin{equation}
\frac{d^2{\hat{\sigma}}^{(0)}_{ab \to t\bar t}}{dt \, du}= F^B_{ab \to t{\bar t}}(c_{tG}) \; \delta(s_4)\, ,
\label{LO}
\end{equation}
where $F^B_{ab \to t{\bar t}}(c_{tG})$ is the squared Born matrix element divided by $16\pi s^2$ and is a quadratic function of $c_{tG}$. The explicit expressions of $F^B_{ab\to t{\bar t}}$ for SM and SMEFT contributions and for the different partonic channels can be found in~\cite{Kidonakis:2023htm}.

\section{Soft-gluon corrections for $t \bar t$ production}
We consider $t{\bar t}$ production with SM and SMEFT partonic processes  as in Eq. (\ref{processes}). When considering soft-gluon corrections in single-particle-inclusive kinematics, 
we use the partonic threshold variable $s_4$ introduced below Eq. (\ref{processes}) which may equivalently be written as $s_4=(p_{\bar t}+p_g)^2-m_t^2$, where $p_g$ is the momentum of an additional gluon in the final state. Near partonic threshold for the production of the $t{\bar t}$ pair, we have $p_g \to 0$ and, thus, $s_4 \to 0$. 
The soft-gluon corrections in the perturbative expansion involve $[(\ln^k(s_4/m_t^2))/s_4]_+$, where $0 \le k \le 2n-1$ at $n$th order in $\alpha_s$.
 
The resummation of soft-gluon corrections follows from the factorization properties of the $t{\bar t}$ cross section under Laplace transforms. 
The resummed cross section is 
\begin{eqnarray}
d{\tilde{\hat \sigma}}_{ab \to t{\bar t}}^{\rm resum}(N,\mu_F,c_{tG}) &=&
\exp\left[\sum_{i=a,b} E_{i}(N_i)\right] \, 
\exp\left[\sum_{i=a,b} 2 \int_{\mu_F}^{\sqrt{s}} \frac{d\mu}{\mu} \gamma_{i/i}(N_i)\right] 
\nonumber \\ && \hspace{-2mm} 
\times {\rm tr} \left\{H_{ab \to t{\bar t}}\left(\alpha_s(\sqrt{s}),c_{tG}\right) {\bar P} \, \exp \left[\int_{\sqrt{s}}^{{\sqrt{s}}/N} \frac{d\mu}{\mu} \, \Gamma_{\! S \, ab \to t{\bar t}}^{\dagger} \left(\alpha_s(\mu)\right)\right] \right.
\nonumber \\ && \hspace{6mm}
\left. 
\times {\tilde S}_{ab \to t{\bar t}} \left(\alpha_s\left(\frac{\sqrt{s}}{N}\right)\right) \;
P\, \exp \left[\int_{\sqrt{s}}^{{\sqrt{s}}/N} \frac{d\mu}{\mu} \, \Gamma_{\! S \, ab \to t{\bar t}} \left(\alpha_s(\mu)\right)\right] \right\} 
\label{resummed}
\end{eqnarray}
where the first exponent, $E_i$, resums collinear and soft contributions from the incoming partons, and the second exponential involves the factorization-scale dependence in terms of the anomalous dimension, $\gamma_{i/i}$, of the parton density. The resummation of noncollinear soft-gluon emission is derived via the soft anomalous dimension matrices $\Gamma_{\! S \, q{\bar q} \to t{\bar t}}$ and $\Gamma_{\! S \, gg \to t{\bar t}}$ \cite{NKGS,NK2l}. For additional  details and references see Ref.~\cite{Kidonakis:2023htm}.

\section{Results at LHC energies}
In this section we present numerical results for the total $t{\bar t}$ cross section and the top-quark differential distribution at LHC energies. We calculate the effect of the dimension-six operator in Eq.~(\ref{eftop}) including diagrams containing at most one effective operator insertion. 

\begin{figure}[htbp]
\begin{center}
\includegraphics[width=72mm]{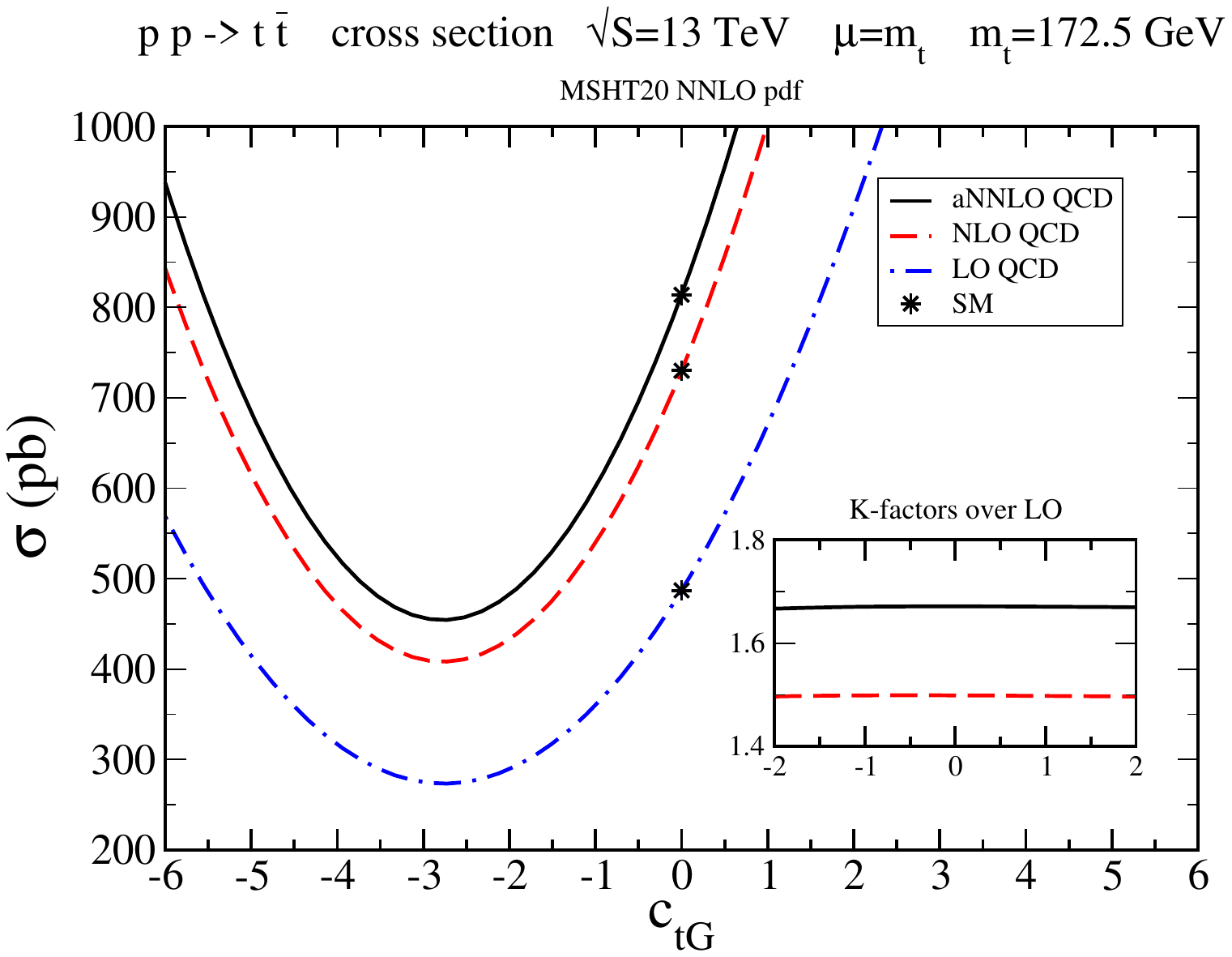}
\includegraphics[width=72mm]{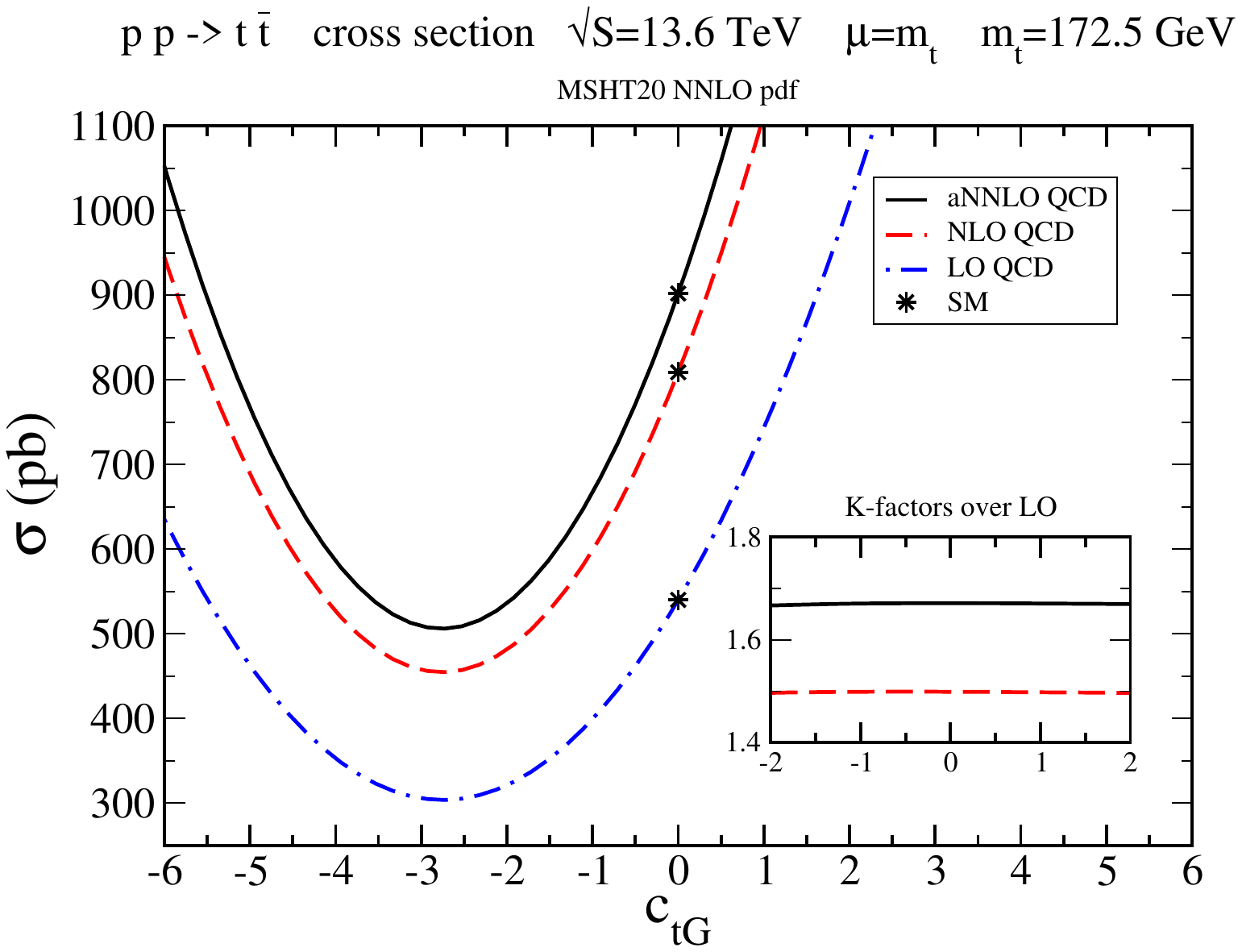}
\caption{LO (blue dot-dashed), NLO (red dashed), and aNNLO (black solid) total cross section for $t{\bar t}$ production in $pp$ collisions at LHC 13 TeV (left) and 13.6 TeV (right), using MSHT20 NNLO pdf \cite{MSHT20nnlo}, as a function of $c_{tG}$. The black stars highlight the SM prediction of the cross section. The inset plots show the NLO over LO and aNNLO over LO $K$-factors.}
\label{xseckf}
\end{center}
\end{figure}

The cross section is a polynomial of second degree in $c_{tG}$: 
\begin{equation}\label{sigma_smeft}
\sigma(c_{tG})=\beta_0 + \frac{c_{tG}}{(\Lambda/1 {\rm TeV})^2} \beta_1
+\frac{c_{tG}^2}{(\Lambda/1 {\rm TeV})^4} \beta_2 \, ,
\end{equation}
where $\beta_0$ is the SM cross section, $\beta_1$ is the contribution from the interference of SM diagrams and diagrams with one SMEFT insertion, and $\beta_2$ is derived by squaring diagrams with one SMEFT insertion. 
Results at 13 TeV and 13.6 TeV are summarized in Fig.~\ref{xseckf}.

\begin{figure}[htbp]
\begin{center}
\includegraphics[width=72mm]{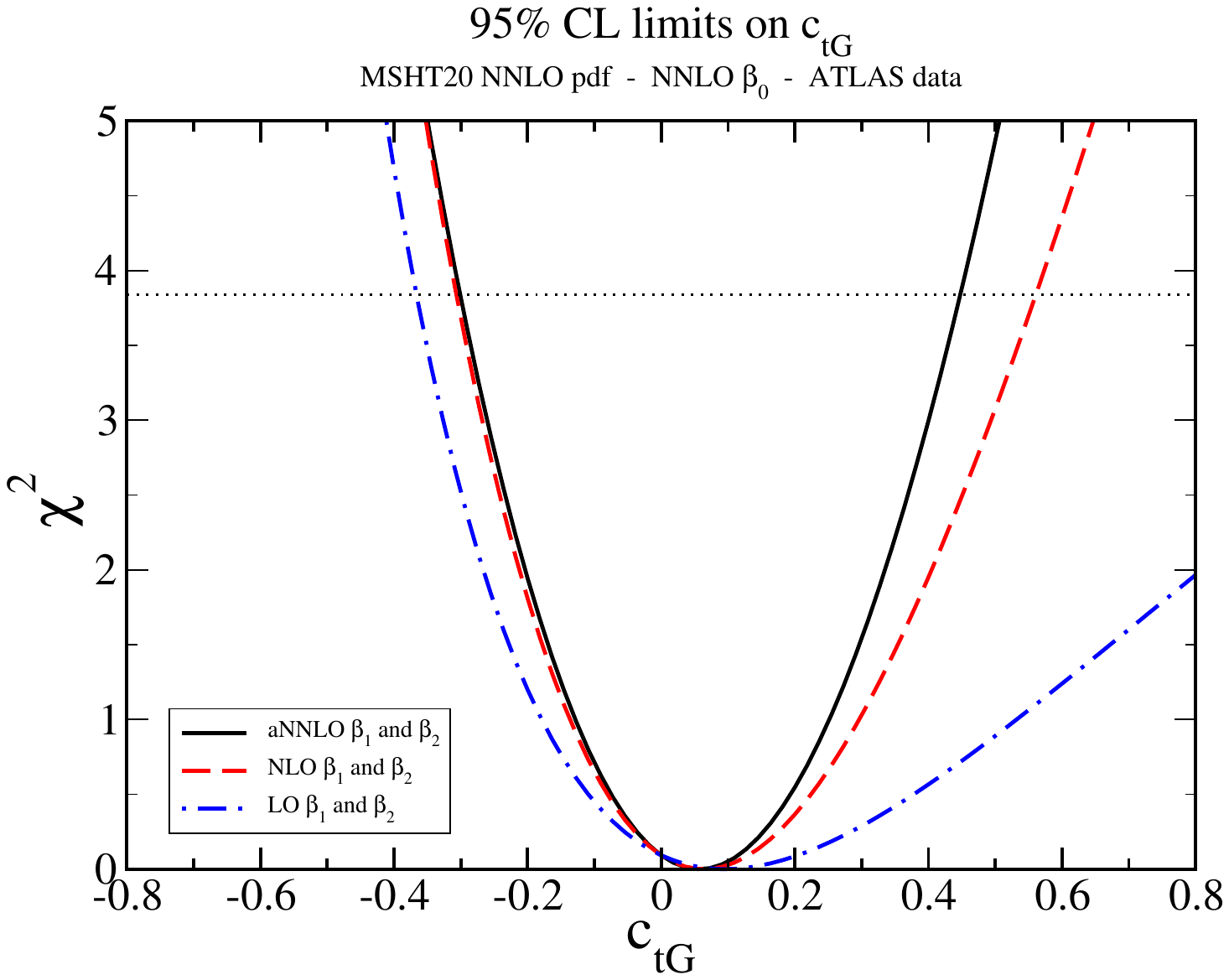}
\includegraphics[width=72mm]{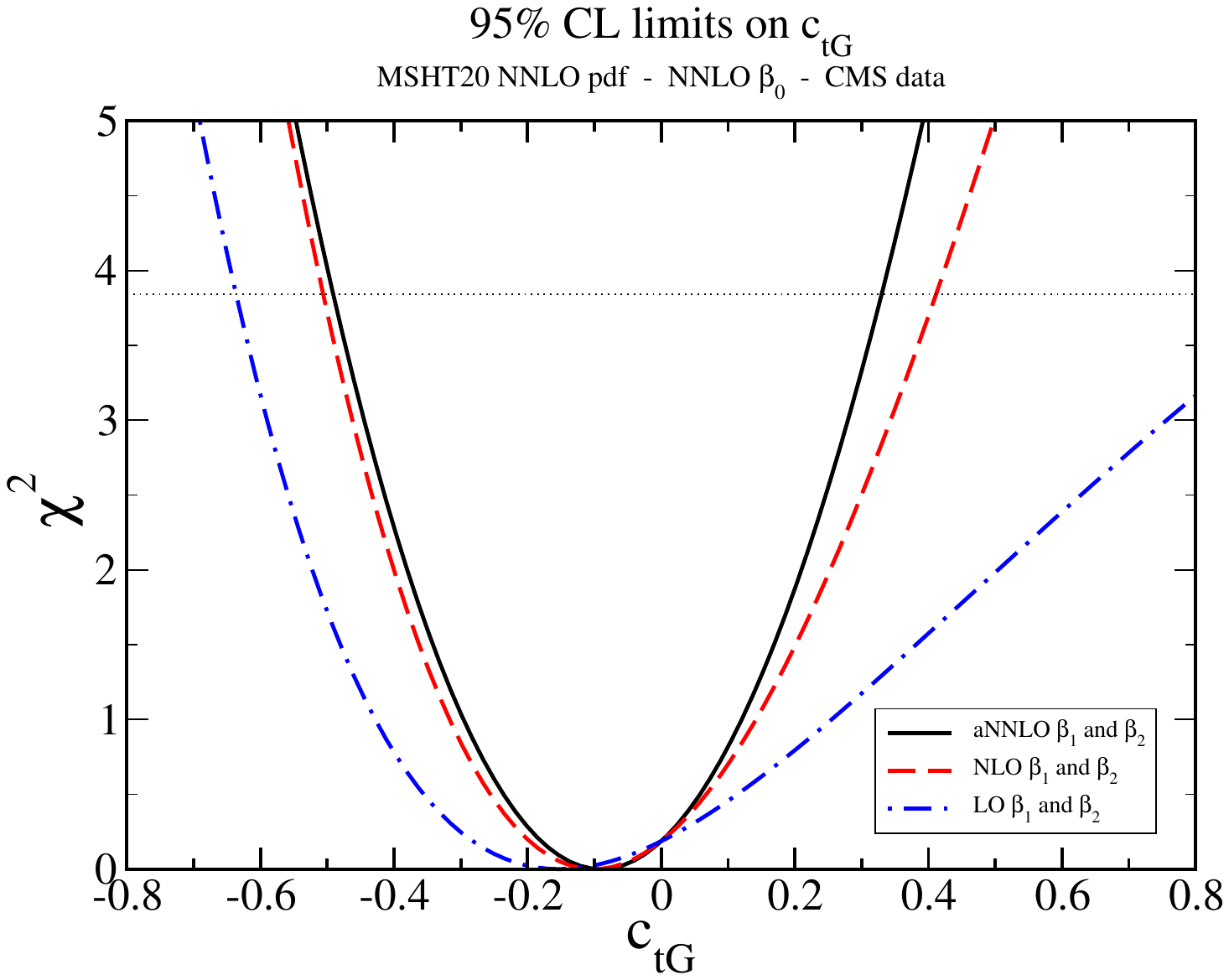}
\caption{Chi-squared as function of the effective operator coefficient $c_{tG}$ constructed using LO (blue dot-dashed curve), NLO (red dashed curve), and aNNLO (black solid curve) SMEFT contributions $\beta_1$ and $\beta_2$. The left and right plots use, respectively, the ATLAS \cite{ATLAS:2023gsl} and CMS \cite{CMS:2021vhb} measurements as experimental inputs.}
\label{chi2atlas}
\end{center}
\end{figure}

To estimate the impact of our aNNLO QCD calculations on a SMEFT global fit analysis, we compare our theoretical predictions with the most precise measurements of the total cross section of top-quark pair production at 13 TeV to set limits on the $c_{tG}$ coupling. We take as experimental input the latest ATLAS result of $829\pm 15$ pb~\cite{ATLAS:2023gsl} and CMS result of $791 \pm 25$ pb~\cite{CMS:2021vhb}. To find the 95\% CL exclusion limits, we set $\Lambda=1$ TeV and construct the chi-squared function
\begin{equation}\label{chi2f}
\chi^2(c_{tG})=\frac{[\sigma_{\rm exp}-\sigma(c_{tG})]^2}{\delta\sigma_{\rm exp}^2+\delta\sigma(c_{tG})^2} \, , 
\end{equation}
where $\sigma_{\rm exp}$ is one of the experimental measurements of the total $t\bar t$ cross section with uncertainty $\delta\sigma_{\rm exp}$, and $\sigma(c_{tG})$ is the theoretical cross section as a function of $c_{tG}$. In Fig.~\ref{chi2atlas} we derive 95\% CL limits on $c_{tG}$ using ATLAS (left plot) and CMS (right plot) results. We show three curves representing the chi-squared function of Eq.~(\ref{chi2f}) obtained by using LO (blue dot-dashed line), NLO (red dashed line) and aNNLO (black solid line) QCD predictions for the SMEFT contributions $\beta_1$ and $\beta_2$. In both figures, we use as SM contribution $\beta_0$ the NNLO QCD result.

In addition to the total $t{\bar t}$ cross sections, we have computed SM and SMEFT contributions to the top-quark $p_T$ distribution through aNNLO in QCD, in six $p_T$ bins up to 550 GeV. The corrections at NLO are large and the additional enhancements at aNNLO are significant. Further details are provided in Ref.~\cite{Kidonakis:2023htm}. 

\section{Conclusions}
We have calculated higher-order corrections to $t{\bar t}$ production at the LHC in SMEFT. In particular, we have added soft-gluon corrections at aNNLO to the exact NLO result in the presence of the chromomagnetic dipole operator. The SM and SMEFT contributions to the total cross sections receive similar enhancements (50\%) from the NLO QCD corrections at 13 and 13.6 TeV LHC energies. The aNNLO QCD corrections are significant, contributing a further 17\% enhancement of the cross section for both the SM and SMEFT cases, and they substantially reduce the scale variation of the cross section. 

Comparing our theoretical predictions with the latest ATLAS~\cite{ATLAS:2023gsl} and CMS~\cite{CMS:2021vhb} results, we have found that the aNNLO corrections improve the lower bound on the $c_{tG}$ coefficient by 2 to 5\%, and they reduce the upper bound by 23 to 35\%, where the specific numbers depend on the experimental input and the SM prediction. Thus, the aNNLO contributions are important for improving the sensitivity on SMEFT operators in global-fit analyses.

We have also calculated top-quark transverse-momentum distributions in the SM and SMEFT. In the case of the $p_T$ distribution, our results show that the binned SMEFT and SM $K$-factors present sizable differences, contrary to the total cross section case, where we have instead similarities between SMEFT and SM $K$-factors (additional details are provided in Ref.~\cite{Kidonakis:2023htm}). The top-quark $p_T$ distribution has been used as an observable in experimental studies~\cite{CMS:2018adi,ATLAS:2022xfj,ATLAS:2022mlu} of top-pair production in order to set limits on the chromomagnetic dipole operator. Therefore, our calculation through aNNLO of the modifications induced by the chromomagnetic dipole operator to the top $p_T$ distribution is relevant when performing fits to improve the sensitivity to this operator in SMEFT searches.

\section*{Acknowledgements}
This material is based upon work supported by the National Science Foundation under Grant Nos. PHY 2112025 and PHY 2412071.

\end{document}